\documentclass[final,3p,times,twocolumn,fixfloat]{elsarticle}
\usepackage{graphicx}
\usepackage{amssymb} 
\usepackage{amsmath} 
\usepackage{tabularx}

\usepackage{lineno}

 \usepackage{ifpdf}
 \ifpdf
 \usepackage[pdftex]{hyperref}
 \else
 \usepackage[hypertex]{hyperref}
 \fi

 \hypersetup{
   pdftitle={},%
   pdfauthor={},%
   pdfsubject={},%
   pdfkeywords={},%
   pdfstartview={},%
   bookmarksopen=true, breaklinks=true, debug=true, %
   colorlinks=true, linkcolor=red, citecolor=blue, urlcolor=blue
 }

\newcommand{\glabcms}{\gamma^{\rm lab}_{\rm c.m.s.}}

\newcommand{\dylabcms}{\Delta y^{\rm lab}_{\rm c.m.s.}}

\journal{Advances in High Energy Physics}

\usepackage{float} 
\usepackage{subfig} 

\usepackage{amsmath}
\usepackage{epsfig}
\usepackage{float}
\usepackage{verbatim} 
\usepackage{color}

\newcommand{\ie}{{\it i.e.}}
\newcommand{\eg}{{\it e.g.}}

\newcommand{\cf}[1]{{Fig.~\ref{#1}}}

\newcommand{\ct}[1]{{Table~(\ref{#1})}}

\begin{document} 
\begin{frontmatter}

\title{Near-threshold production of $W^\pm$, $Z^0$ and $H^0$ at a fixed-target experiment at the future ultra-high-energy proton colliders}

\author[IPNO]{J.P.~Lansberg}
\ead{lansberg@in2p3.fr}
\author[IPNO,AAU]{R. Mikkelsen}
\ead{rune@phys.au.dk}
\author[AAU]{U.I.~Uggerh{\o}j}
\ead{ulrik@phys.au.dk}
\address[IPNO]{IPNO, Universit\'e Paris-Sud, CNRS/IN2P3, F-91406, Orsay, France}
\address[AAU]{Department of Physics and Astronomy, Aarhus University, 8000 Aarhus, Denmark}

\begin{abstract}
\small

We outline the opportunities to study the production of the Standard Model bosons, 
$W^\pm$, $Z^0$ and $H^0$ at ``low'' energies at fixed-target experiments based at possible future 
ultra-high-energy proton colliders, \ie\ the High-Energy LHC, the Super proton-proton Collider 
and the Future Circular Collider -- hadron-hadron. 
These can be indeed  made in conjunction with 
the proposed future colliders designed to reach up to $\sqrt{s}=100$ TeV by using bent 
crystals to extract part of the halo of the beam which would then impinge on a fixed 
target. Without disturbing the collider operation, this technique allows for the 
extraction of a substantial amount of particles in addition to serve for a beam-cleaning 
purpose. With this method, high-luminosity fixed-target studies at centre-of-mass energies 
above the $W^\pm$, $Z^0$ and $H^0$ masses, $\sqrt{s} \simeq 170-300$ GeV, are possible. 
We also discuss the possibility offered by an internal gas target, which can also 
be used as luminosity monitor by studying the beam transverse shape.

\end{abstract}


\end{frontmatter}




\section{Introduction}

In this Article, we consider the possibility of performing fixed-target experiments 
with the beams of the proposed high-energy LHC (HE-LHC) (see \eg\ \cite{HELHC:2011}), 
the Future Circular Collider hadron-hadron (FCC-hh) (see \eg\ \cite{FCC-hh:2014}) and 
the Super proton-proton collider (SppC)~(see \eg\ \cite{SppC:2013}), which,
as their names indicate, are primarily intended for collider physics. The 
beam energy of these possible future facilities ranges
from 16.5 up to 50 TeV, allowing for fixed-target collisions at centre-of-mass system (c.m.s.) energies
ranging from 175 to 300 GeV.

Just as lower energy beams, these can in principle be extracted over the course 
of about a meter by using the channelling of particles in a bent-crystal.
This phenomenon is well documented (see \eg\ \cite{Elsener:1995hm,Arduini:1997nb,Baurichter:2000wk,Biryukov:2005gr,Scandale:2011zz})  and has been experimentally studied  for protons 
and heavy ions at energies per nucleon up to $900$ GeV.
Recently, studies performed at SLAC have shown that the beam bending by means of bent 
crystals is also possible for high-energy positrons and electrons~\cite{Wienands:2015hda}.
In Ref.~\cite{Uggerhoj:2005xz}, it was discussed specifically for the LHC beams.  
Channelling experiments at the LHC have been proposed~\cite{LUA9}, 
are installed~\cite{LHCC107,LMC-173} and will be performed during the Run-2 for beam 
collimation studies. The bent-crystal extraction technique allows for the extraction of 
 particles  from the beam halo only, so that the 
collider experiments can be kept running simultaneously. These particles would
anyway be lost to collimation and would not be used in the collider mode.

In Ref.~\cite{Brodsky:2012vg}, a comprehensive list of physics opportunities offered by the use of
 the multi-TeV proton and lead LHC beams on a fixed target was presented. 
Let us recall the critical assets of the fixed-target mode
as compared to the collider mode \ie\
\begin{itemize} 
\item[$\cdot$] a quasi unlimited target-species versatility,
\item[$\cdot$] a full access to the target rapidity, which corresponds to the far backward region in the centre-of-mass frame,
\item[$\cdot$] the possibility to polarise the target,
\item[$\cdot$] very large luminosities with modest beam intensity thanks to the high target density.
\end{itemize}

These advantages already translate, with the proton and lead LHC beams, into an impressive list 
of possible physics studies~\cite{Brodsky:2012vg} beyond their respective state-of-the-art, in
particular as regards precision studies of spin-related observables with a polarised target, 
of quark-gluon  plasma formation in lead-nucleus collisions in the target rapidity region 
at c.m.s. energies between those of SPS and RHIC 
 and of QCD at large momentum fractions $x$ in proton-proton, 
proton-deuteron and proton-nucleus collisions, etc. We refer to Ref.~\cite{Lansberg:2012kf} for the 
more specific case of quarkonium studies, for spin physics to 
Refs.~\cite{Lorce:2012rn,Rakotozafindrabe:2013au,Lansberg:2014myg,Massacrier:2015nsm} and for 
heavy-ion physics to Refs.~\cite{Rakotozafindrabe:2012ei,Lansberg:2012kf,Lansberg:2013wpx}. 
First simulation studies at the generator level have been presented in Ref.~\cite{Massacrier:2015qba} 
and have demonstrated the great potential for both charmonium and bottomonium studies at $\sqrt{s}=72$ 
and $115$~GeV at a fixed-target experiment with the LHC beams (thereafter referred to as AFTER@LHC).


With beams of higher energies at future facilities, the available c.m.s. energies can  nearly 
be three times as large as at AFTER@LHC and allow for even more systematic studies of systems whose masses 
are well above that of the bottomonia, \ie~10~GeV. At fixed-target LHC energies, $W$ and $Z$ 
production, sometimes generically referred to as the Drell-Yan-like processes, is just reachable 
with very low expected rates but with the advantage of potentially providing unique 
information about the nucleon structure at momentum fractions $x$ close to unity ($x\simeq M/\sqrt{s} e^{\pm y_{c.m.s.}}$), about QCD 
corrections near the threshold\footnote{Let us note here the production of heavy 
Beyond-the-Standard-Model particles produced at the LHC in the collider mode might also 
subject to similar QCD threshold corrections.} and of offering interesting information on hadronic 
$W$ decays. With beams of higher energies, rates would significantly be larger allowing, 
among other things, for rapidity dependent measurements. In general, 
the combination of high luminosity hadron-hadron collisions at $\sqrt{s}$ well above 
100 GeV and a backward c.m.s coverage provide the opportunity to study the interplay 
between the --genuinely non-perturbative-- confinement of partons at  large momentum 
fractions $x$ that are involved in extremely hard --perturbative-- reactions which are believed 
to be well understood within the perturbative regime of QCD. Using a polarised target 
allows one to advance further the precision and the refinements of such studies of the 
hadron inner structure with information on the helicity of the partons and on their 
angular momentum when they carry most of the hadron momentum. 
The confinement properties of Quantum Chromodynamics, the 
theory of strong interaction, is still an open problem which deserves novel and 
innovative studies, even at high energy facilities.

Although, exactly as for AFTER@LHC,
the potential for physics studies go well beyond that of Drell-Yan-like studies in the mass region 
of $W$ and $Z$ bosons, 
we wish to focus on it here as an illustrative example of the gain offered by even higher energy beams --in what we believe to be the very first article
on the use of ultra-high energy beam in the fixed-target mode.
The case for such $W$ and $Z$ studies is clear. So far, the production of $W^\pm$ and $Z$ bosons at RHIC could 
only be performed at $\sqrt{s}=500$~GeV (see \eg~\cite{Adare:2010xa,Aggarwal:2010vc})
with a couple of thousands $W^\pm$ candidates and less than one hundred $Z$ counts. 
Studies at lower energies, in order to reach $x$ higher than 0.2 and measure the
 $\bar d /\bar u$ flavour asymmetry at larger $x$ and high $Q^2$, require luminosities on the order of an 
inverse femtobarn which is out of reach
at RHIC in particular if operated at $\sqrt{s}=200$~GeV where its luminosity is significantly lower. In addition, the most accessible region in the fixed-target mode at high energies is that of backward c.m.s. 
rapidities where $x$ in the target can even be larger.  
Of course, $H^0$ production has so far 
only been observed at the LHC at 7 and 8 TeV~\cite{Aad:2012tfa,Chatrchyan:2012ufa} and
it is is of interest to have a look at the conditions in which it could be studied not far from the threshold.

The structure of this Article is thus as follows. In section 2, we discuss the beam extraction
in a parasitic mode using the  bent crystal technique at ultra high energies. In section 3, we recall
the main characteristics of possible future colliders as they are currently discussed and derive
reasonnable expectations for luminosities in the fixed-target mode. In section 4, we discuss the case of the 
weak boson production as a benchmark of what can be achieved with luminosities 
up to 100 times larger than at RHIC. In addition to the expected rates, we  briefly discuss
the potential backgrounds whose precise size can however only be assessed with a 
proposed detector set-up. Section 5 gather our conclusions.

\begin{table*}[!ht]
\begin{center}\renewcommand{\arraystretch}{1.3}
    \begin{tabularx}{17cm}{ p{5cm} | c | c |c |c }
     & SppC-1 & SppC-2 & HE LHC & FCC-hh  \\ \hline
     Beam Energy ($E_p$) [TeV] & 20 & 45 & 16.5 & 50 \\ 
     Fixed-target centre-of-mass energy  ($\sqrt{2E_p m_N}$) [GeV] & 194 & 291 & 176 & 307\\ 
     Number of bunches stored  & 3000 & 6000 & 1404 (50 ns spacing) & 10600/53000 (25 and 5 ns spacing) \\ 
     Number of protons ($N_p$) per bunch [$10^{11}$] & 1.7$\cdot 10^{-3}$ & 0.98$\cdot 10^{-3}$ & 1.3 & 1/0.2  \\ 
$\glabcms=\frac{\sqrt{s}}{2m_p}$ &  103  & 155   & 94  & 163   \\
$\dylabcms =\ln(\glabcms+\sqrt{(\glabcms)^2-1})$ &  5.3  & 5.7   & 5.2  & 5.8   \\
    \end{tabularx}
\end{center}
\caption{Beam parameters for the proposed next-generation colliders and the 
corresponding fixed-target energies along with  the boost ($\glabcms$)  
and the rapidity shift ($\dylabcms$) between the centre-of-mass frame of the 
fixed-target collision and the laboratory frame -- which identifies to the centre-of-mass frame
in the collider mode.}
              \label{table:BeamParameters}
\end{table*}

\section{Beam extraction by means of a bent crystal with ultra high energy protons}
As aforementioned, these possible future collider facilities would use
proton beams from 16.5 up to 50 TeV. The bending of GeV beams of protons and ions 
has been studied extensively during the past three decades. As a first approximation, 
one may calculate the approximate deflection efficiency as a function of crystal 
length as \eg\ done in Ref.~\cite{Baurichter:2000wk}. For example, at a deflection 
angle of 0.5 mrad, as approximately required for the passage of a septum blade 
downstream required for further extraction, the efficiency (excluding surface 
transmission) in Si (110) is 84\% for a 50~TeV beam. This efficiency is obtained 
at the optimum crystal length of $L/L_D=0.085$ (see \cf{fig:Efficiency}), 
corresponding to a length of $1.6$~m.

\begin{figure}[htbp]
\begin{center}
\includegraphics[width=0.475\textwidth]{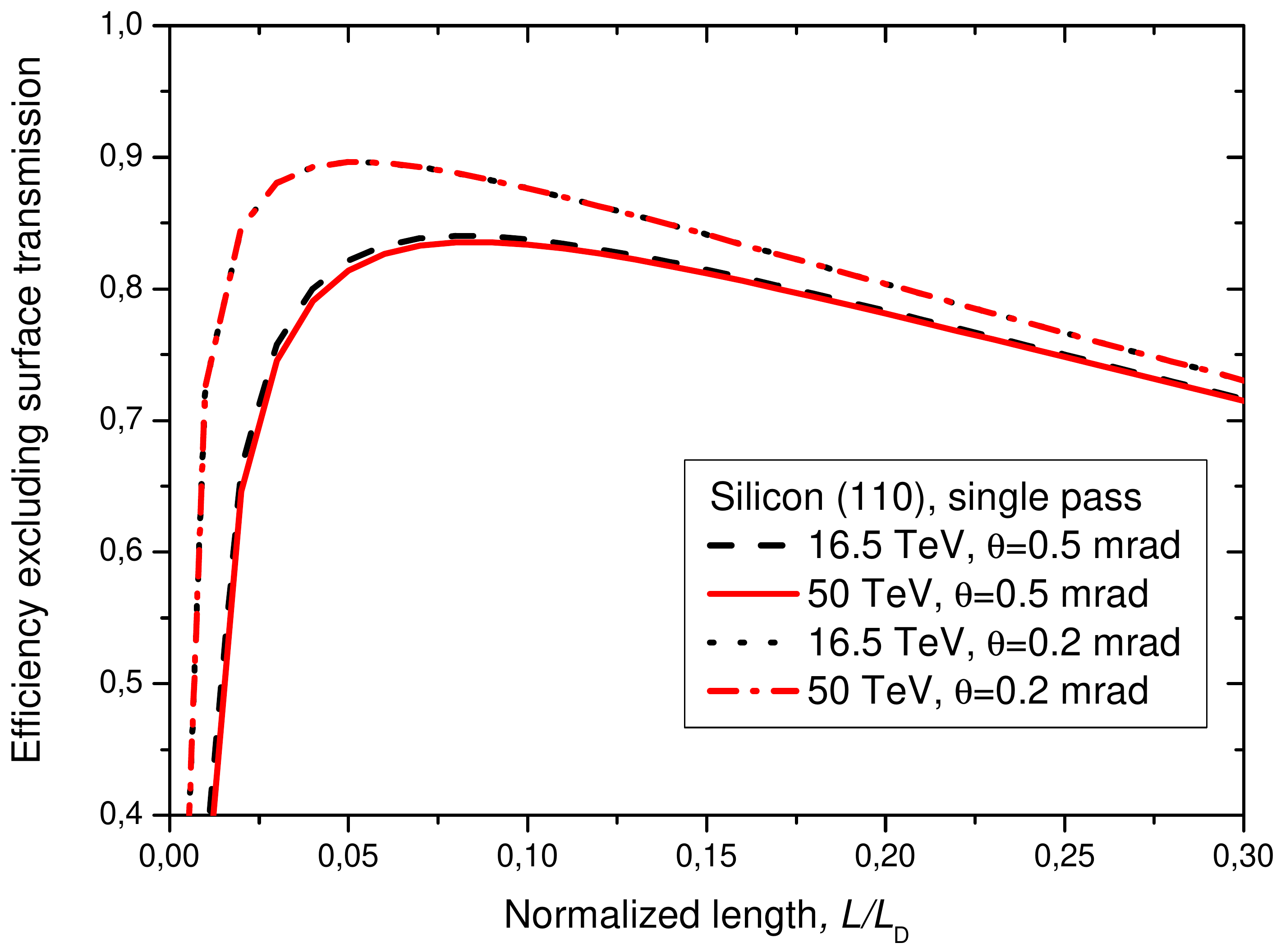}
\caption{Single-pass efficiency, excluding surface transmission, calculated as in Ref.~\cite{Baurichter:2000wk}. The angles and energies are given in the legends.}
\label{fig:Efficiency}
\end{center}
\end{figure}

However, the optimal distance calculated in this simulation may significantly be overestimated 
because it does not take into account the probability for dechanneled particles 
to get extracted on a later encounter with the crystal. Several studies have shown 
that this multi-pass mechanism may result in a significant shortening of the optimal distance. 
The optimal crystal thickness also depends on the beam optics used.
Results of the order $20-30$ cm are certainly not unrealistic.

\begin{table*}[!ht]
\begin{center}\renewcommand{\arraystretch}{1.2}
    \begin{tabular}{ c | c  | c | c |c }
     & SppC-1 & SppC-2 & HE LHC & FCC-hh  \\ \hline
Extracted beam &&&&\\
     Proton flux & $7.1\cdot10^{5}$ & $8.1\cdot10^{5}$ & $2.5\cdot10^8$ & $1.5\cdot  10^9$ \\ 
     $\mathcal{L} (\mu $b$^{-1}$s$^{-1})$ & 0.028/0.088/0.044 & 0.032/0.10/0.05 & 10/31/15 & 30/93/45 \\ 
     $\int dt \mathcal{L}($pb$^{-1}$yr$^{-1})$ & 0.28 / 0.88 / 0.44 & 0.32/1.0/0.5 & 100/310/155 & 300/930/450 \\ 
   \hline
   Gas target &&&&\\
$\mathcal{L}(\mu $b$^{-1}s^{-1})$& 0.014 &0.016&5&30\\
     $\int dt \mathcal{L}($pb$^{-1}$yr$^{-1})$ & 0.14 & 0.16  & 50 & 300 
        \end{tabular}
\end{center}
             \caption{The proton flux is calculated by assuming that 5 \% of the beam is used per fill of 10 hours. The luminosities are calculated for the case of targets that are 1 cm thick. The three values displayed represent luminosities for three different targets: liquid Hydrogen, Beryllium and Tungsten. The gas-target values are calculated using the same parameters as in~\cite{Massacrier:2015nsm} for a perfect gas at a pressure of $10^{-9}$ Bar in a zone of 100 cm.}
              \label{table:extracted}
\end{table*}

\section{Main parameters of future colliders and their corresponding characteristics in the fixed-target mode}
\label{sec:FuturColl}

In order to derive the luminosities which can reasonably be expected in the 
fixed-target mode with the beams of future colliders, we start by recalling their main 
parameters as currently discussed. Indeed, 
efforts are now being made in investigating the best location and technology for future 
collider projects. These proposed circular colliders have circumferences of $50$ 
to $100$ km and the ability to circulate protons with $15$ to $50$ TeV energies. 
We list their most relevant characteristics in 
\ct{table:BeamParameters}. In particular, we consider the phase one and two of
 the SppC, denoted SppC-1 and SppC-2, as discussed in Ref.~\cite{SppC:2013}. We also
consider the High Energy Large Hadron Collider (HE LHC)~\cite{HELHC:2011} and the Future
 Circular Collider hadron-hadron (FCC-hh)~\cite{FCC-hh:2014}.

With colliding beams of equal energies, the c.m.s. frame obviously corresponds to
the laboratory frame. In the fixed-target mode, with the LHC 7 TeV protons for instance, 
the boost ($\glabcms$)   and the rapidity shift ($\dylabcms$) between the c.m.s. frame of the fixed-target collision and the laboratory frame are respectively
$\glabcms=\sqrt{s}/(2m_p)\simeq 60$ and $\dylabcms =\ln(\glabcms+\sqrt{(\glabcms)^2-1}) \simeq 4.8$. The region of central c.m.s. rapidities, $y_{\rm c.m.s.}\simeq 0$, 
is thus highly boosted at an angle with respect to the beam axis of about one degree 
in the laboratory frame. The entire backward hemisphere, $y_{\rm c.m.s.}<0$, is thus easily accessible with standard experimental techniques. With the future facilities, the rapidity shift is on the order of 5-6, see \ct{table:BeamParameters}. A detector covering 
$\eta_{\rm lab} \in [2,6]$ would thus cover nearly half
of the physical phase space of the fixed-target mode.

As we discussed in the previous section, the extraction of such high energy beams 
by a bent crystal should not 
pose more 
challenges than at the LHC where it will be tested in the coming year. In this case, 
the main accelerator parameter fixing the luminosities achievable is
the flux of the extracted beam. In the following discussion, we will assume\footnote{See~\cite{Brodsky:2012vg} for a discussion on the LHC conditions where it 
 corresponds to half of the beam loss.} that
it amounts to 5\% of the protons stored in the beam over a fill lasting 10 hours. 
In the case of the LHC, such a parasitic mode
corresponds to a proton flux of $5 \times 10^8$ per second and, on the average, 
to the extraction of mini-bunches of about 15 protons per bunch per pass with a 
25 ns bunch spacing. In such a case, 
with a target thickness of 5-10 \% of interaction length --which is the case we consider 
here--, the pile-up is not an issue. The corresponding numbers for future facilities 
are given in \ct{table:extracted}.

\begin{figure*}[htbp]
\begin{center}
\includegraphics[width=0.7\textwidth]{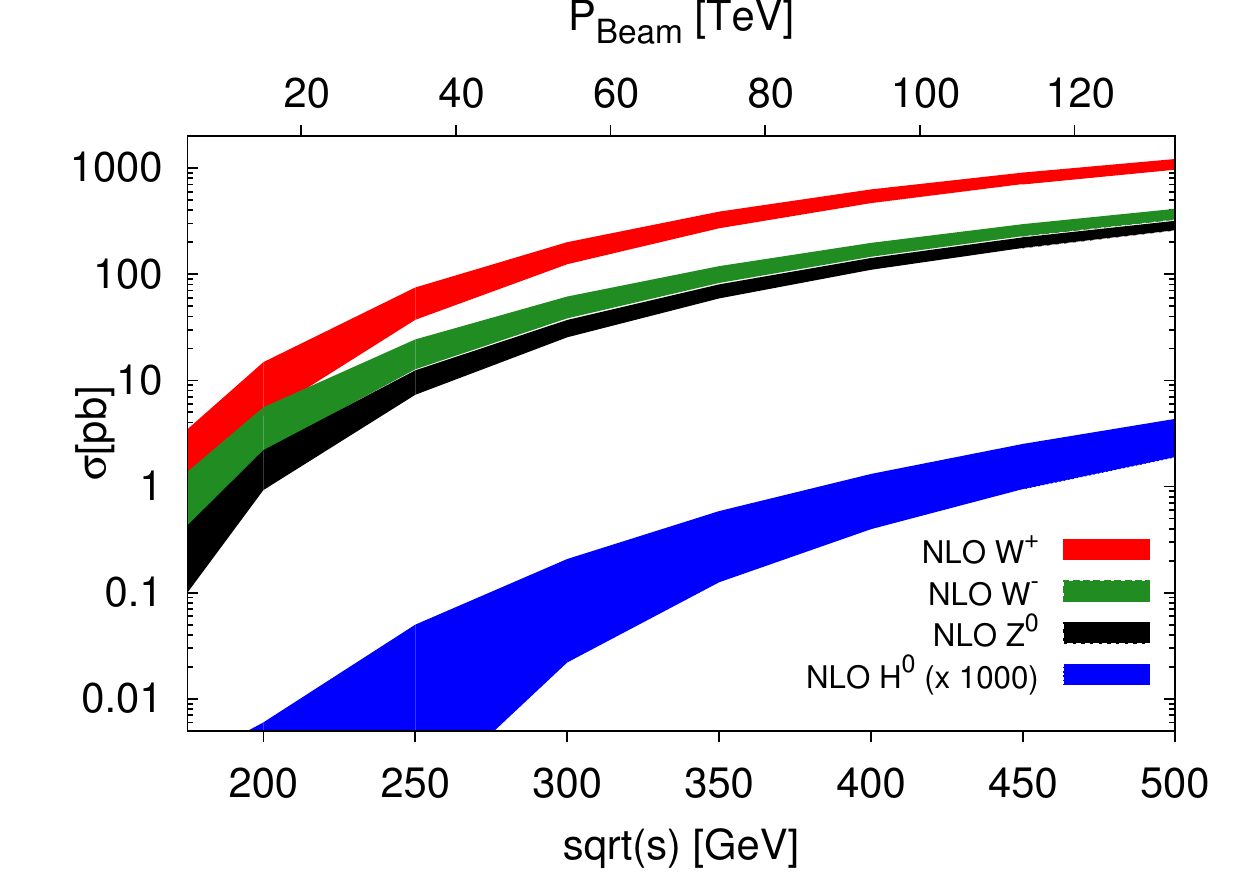}
\caption{Cross sections for Standard Model boson production in proton-proton collisions at various centre-of-mass (lower $x$ axis) and corresponding beam energies in the fixed-target mode  (upper $x$ axis). The color bands indicate the 1-sigma values coming from the PDF uncertainty. The NLO calculations were performed with CT10NLO~\cite{Lai:2010vv} by using {\sc mcfm}~\cite{Campbell:2000bg} with $\mu_F=\mu_F$ set to the mass of the produced particle. The 4 orange arrows point at the beam energy of the 4 set-ups which we have considered.}
\label{fig:NLO}
\end{center}
\end{figure*}

 Yet, it may not 
be necessary to extract the beam from the collider to perform fixed-target experiments. 
By injecting a small amount of gas into the detector region of a running collider, 
one can sufficiently increase the probability of beam-gas interactions 
such as to reach decent
luminosities yet without affecting at all the beam lifetime.
At the LHC, the LHCb experiment has implemented such a system initially to monitor 
the beam luminosity~\cite{LHCb:SMOG,FerroLuzzi:2005em,LHCb:2014} referred to as 
SMOG for  System for Measuring Overlap with Gas (SMOG). SMOG has so far proved 
to be functioning well while not disturbing the primary beam. LHCb is currently 
analysing data in proton-neon and lead-neon collisions taken during beam tests in 2012 and 2013.
In Ref.~\cite{Massacrier:2015nsm}, the corresponding luminosities are given for the LHC. 
One may  think that switching from a dense solid or liquid target to a dilute 
($10^{-9}$ Bar) gas necessarily decreases the luminosity. In fact, it may not be always so.
Indeed, this decrease is  compensated since the entire collider beam, amounting to a current 
close to an ampere for the LHC, traverses the gas cell, as opposed to the
extraction beam which is similar to the beam loss. 

\ct{table:extracted} summarises all these numbers for 1cm thick targets in the case of an 
extracted beam with a bent crystal and an internal gas target of 1 meter.
In the former case and for light target materials, the luminosity can be increased by using a  
target much thicker/longer than 1 cm -- NA51 at SPS used for instance a $1.2$ m long hydrogen  
and deuterium target~\cite{Baldit:1994jk} with 450 GeV protons, E866 at Fermilab used a 
3 target cells of 50 cm~\cite{Zhu:2007aa} with 900 GeV protons, COMPASS at SPS 
uses~\cite{Gautheron:2010wva,Chiosso:2013ila} a 110 cm polarised NH$_3$ target with 160 GeV pions. 
In such a case,  one can obtain luminosities {\it per annum} well above the inverse femtobarn.
Even in this case, the thickness of the target does not reach more than 10 \% of interaction length.

Much higher luminosities could be achieved by using the full 
amount of the remaining protons stored in the beam at the end of each fill. Such a 
``dumping'' mode, which could last an hour without unreasonnably impacting the schedule 
of the machine would of course provide luminosities orders-of-magnitude higher 
--probably up to 3 -- than the ones quoted in \ct{table:extracted}. However, 
this would also be done at the cost of carrying out the experiment 
in a highly activated environment, which may not be feasible in practice, and at the cost of
a significant pile-up. For some specific studies to look for rare events with very large momentum tracks, 
the latter may however not be an unsolvable issue.

\section{Fixed-target mode and boson production}

\subsection{Expected signal rates}

As announced we have decided to focus on the production of SM bosons production as an 
illustrutive example of what the high luminosities reachable with the fixed-target mode 
can allow for. Physicswise, by measuring the production of Standard Model bosons $W^\pm$ 
and $Z^0$, the distribution of quarks and antiquarks at large $x$ can be probed both in the proton 
with a hydrogen target and in the nuclei with nuclear targets. 
In particular, it allows one to determine the $\bar d /\bar u$ flavour asymmetry at large $x$ and large scales.
The study of these reactions
not far from the threshold also allows one to validate the theoretical methods to account for 
the so-called threshold resummation effects (see \cite{Mukherjee:2006uu} for $W$ 
production at RHIC).

To evaluate the cross section at NLO accuracy, we used the library {\sc mcfm} 
\cite{Campbell:2000bg} and set the $\mu_F=\mu_R$ equal to the boson mass. Since we 
are mainly interested in illustrating how such a measurement would help to 
better constrain parton distribution (PDF), we only show the theoretical uncertainty
from these as they are currently determined. To this end, we use the 
NLO PDF set CT10~\cite{Lai:2010vv} and its associated eigenvector sets. 
{\sc mcfm} takes all these into account and provide a 1-$\sigma$ uncertainty which
we have depicted on \cf{fig:NLO}. One observes an increasing PDF uncertainty for 
decreasing energies. With a 20 TeV beam, the uncertainity of $W^+$ production 
is as large as a factor of 3 for a total cross section about 10 pb. 

If one sticks to the conventional leptonic decay channels, the branching is on the order of 
10\%. At 200 GeV, with a detector covering pseudorapities from 2 to 6 -- a detector similar 
to LHCb~\cite{Alves:2008zz} with a slightly more forward coverage for instance-- and imposing the lepton transverse momentum (as well as the missing transverse momentum carried by the neutrino) to be larger than 20 GeV as usually done to cut the background (see later), a quick evaluation shows that the acceptance is on the order of 45 \% for the $W$ as well as for the $Z$, although with a smaller branching ratio on the order of 3 \%. 
The central values for the cross sections times the corresponding leptonic branchings at $\sqrt{s}=200$ GeV in this fiducial volume ($P_{T,\ell,\nu_\ell} > 20$ GeV$, 2 < \eta_{\ell,\nu_\ell} <6)$ are therefore 400, 150, 20 fb respectively for the $W^+$, $W^-$ and $Z$.

Relying on the performance of LHCb for similar studies~\cite{Aaij:2012vn}, the efficiency
including that of the triggering, the tracking and additional selections\footnote{for the $W$: the lepton isolation, a cut on the energy deposit to limit the punch-through, the absence of a second lepton with a minimum $P_T$, a vertex cut to remove the heavy-flavour decays.} is around 40~\% for the $W$ and 67\% for the $Z$. 

With a yearly luminosity of 15 fb$^{-1}$ --using for instance a 50 cm long liquid hydrogen target at a facility 
similar to a FCC-hh, see \ct{table:extracted}, one would expect a couple of thousands of $W^+$ events to be measured at 200 GeV. It would be one order of magnitude more at 300 GeV.

 $Z$ boson production is also at reach at 200 GeV with 200 dimuon events, especially if the rather clean environment 
associated to the rather low centre-of-mass energy allows for the use --and the study--
of hadron decay channels. With a detector such as LHCb, both electron and muon decay channels can be used 
(see \eg~\cite{Aaij:2015vua}).

For the $H^0$, the situation is very different with much smaller rates (on \cf{fig:NLO}, the
cross section for $H^0$ are shown in fb for readability). Since the process is dominated 
by gluon fusion -- we have checked that the vector-boson-fusion contribution is negligible
down to low energies --, the PDF uncertainties are very large. Fixed-target luminosities of 
100 fb$^{-1}$ seems to be needed to be able to have a hope to see a $H^0$ signal even with 
the 50 TeV FCC-hh proton beams.

For the $W$ and $Z$ bosons, with a very successful experiment,  one may however be able 
to access regions more backward than $y_{{\rm c.m.s.},V} = \ln(M_V/\sqrt{s})$ and hence to probe 
the quark distribution in the target for $x$ larger than unity in the nuclear case~\footnote{The yields 
  in proton-nucleus collisions with a 1 cm thick gold or lead target should be similar to that in
proton-proton collsions with a 1 m long liquid hydrogen target.}.

\subsection{Expected background}

At such low energies, very few processes can mimic a dilepton pair of mass around 90 GeV or an isolated
lepton of a transverse momentum around 40 GeV accompanied with a missing transverse energy of a 
similar size. In fact, one expects the main background to be of electroweak origin such as 
the $Z\to \ell \ell$ for the $W \to \ell \nu_\ell$ channel where one lepton from the $Z$ 
is lost. $W \to \tau \nu_\tau$ is also known to sometimes mimic a  
$\mu \nu_\mu$ final state,  although usually at lower $P_T$. These should be tractable with data.
As compared to studies at the LHC or at RHIC, QCD backgrounds are in general expected to be smaller.

\begin{figure}[htbp!]
\begin{center}
\includegraphics[width=\columnwidth]{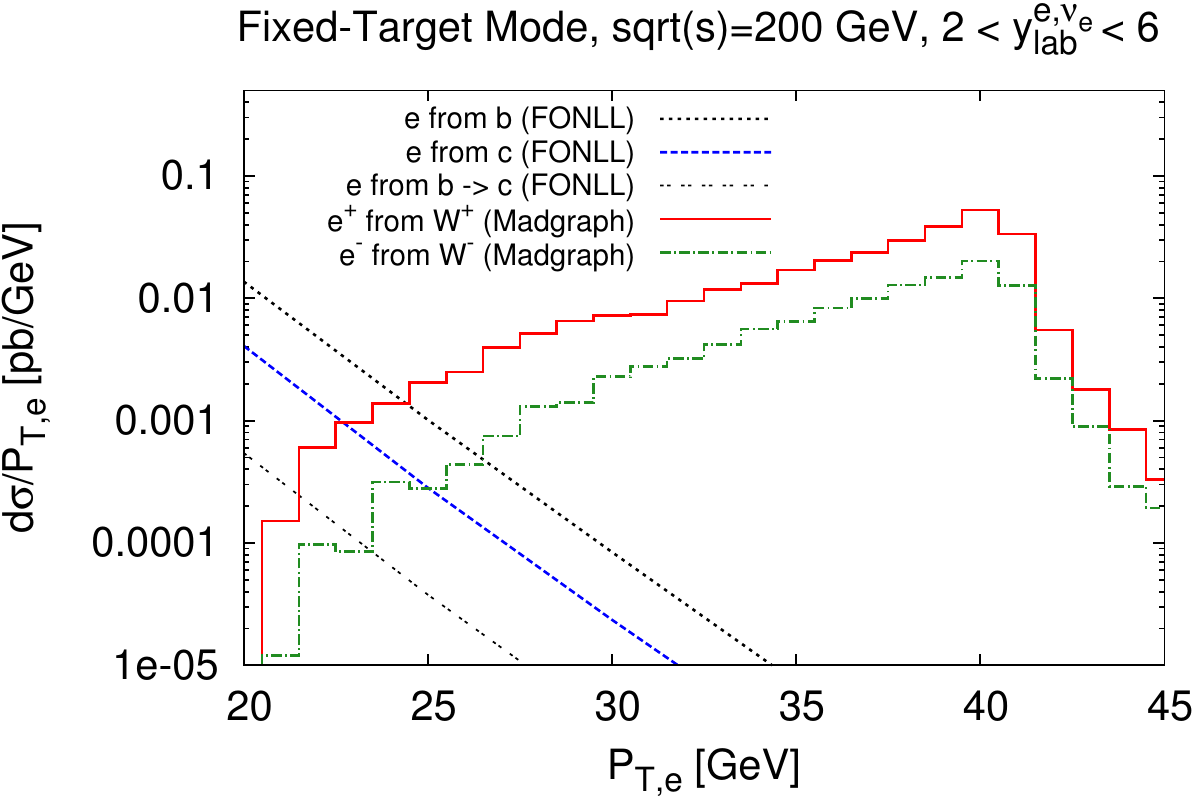}\vspace*{-.25cm}
\caption{$P_T$ spectra for heavy-flavour-decay electrons (and positrons) as predicted by FONLL compared to that from the predicted {\it signal} from $W^\pm$ decays obtained with Madgraph 4~\cite{Alwall:2007st}
at $\sqrt{s}=200$ GeV. The FONLL cross sections have been evaluated with the default set-up of \url{http://www.lpthe.jussieu.fr/~cacciari/fonll/fonllform.html} along the lines of~\cite{FONLL} with the PDFs CTEQ6.6M~\cite{Nadolsky:2008zw}.}
\label{fig:comp-electrons}
\end{center}
\end{figure}

For the $W^\pm$,  \cf{fig:comp-electrons} illustrates that, even without any specific cuts usually used to reduce the background (see below), the $P_T$ spectrum from heavy-flavour-decay electrons is extremely suppressed at large $P_T$ and the electron yield close to $m_W/2$ is essentially
purely from $W$ decays. The same holds for muons. Let us however stress that the main QCD background in the LHCb study of~\cite{Aaij:2012vn} is from hadron decay-in-flight which contributes less than 10 \%. Overall the purity of their $W^\pm$ yield
is 78~\%.

At RHIC, electron channels were used both by STAR~\cite{STAR:2011aa,Adamczyk:2014xyw} 
and PHENIX~\cite{Adare:2010xa,Adare:2015gsd} and they required  its isolation. Since PHENIX has an 
incomplete azimuthal coverage, the requirement for a missing transverse energy could not be imposed. 
Yet, the signal could be extracted. Once these requirements are applied, the $W$ peak in the electron
$P_T$ spectrum around 40 GeV is evident. Let us note here the background electron spectra reported 
in \cf{fig:comp-electrons} would severely be reduced once cuts for a missing transverse energy, for the
isolation of the electron and for a maximum distance from the primary vertex are imposed. Conversely, 
these could be measured --and then subtracted-- by selecting displaced leptons.

With muons, it is even possible not to impose any lepton isolation to extract the $W$ signal
as demonstrated by the CMS study~\cite{Chatrchyan:2012nt} in the very busy lead-lead environment 
at 2.76 TeV. 
At lower $P_T$, QCD backgrounds contribute --essentially
converted photons from $\pi^0$ and $\eta$ decays, whereas muons from converted photon are usually 
negligible. 
The $P_T$ spectrum of this background can be evaluated from dedicated simulations at the detector level 
but its size is however 
very difficult to predict without a precise knowledge of the hadron detector response. Usually
the normalisation of such a background is simply adjusted on the data. In any case, it always 
has been found to be smaller than the signal at electron $P_T$ above 30 GeV. We also note that the
STAR detector is rather slow as compared to the LHCb detector for instance and tracks of particles produced 
in earlier collisions can pile up in the data acquisition system, which further increases the background.

In the $Z$ case~\cite{Aaij:2012vn}, the main background at LHC energies is that of heavy-flavour decays 
with on average 3 background events per thousand ! At lower energies, it should also be even more 
suppressed~\cite{Adare:2010xa}. At 200 GeV for instance, $\sigma_{b \bar b}(2 < \eta_b < 6; P_{T,b}> 20\ {\rm GeV}, m_{b \bar b} > 80\ {\rm GeV}) \sim 220$~fb. Even if one neglects the momentum difference between the $b$ quark, the beauty hadron and the lepton --which is however a far too conservative approximation--, one should multiply it by $f(B\to \ell) \simeq 0.1$ squared. The dilepton background from $b$ decays is thus in any case much smaller than 2 fb whereas the signal size --accounting for the acceptance-- is 20 fb. As regards the uncorrelated background from hadrons, no same-sign dimuon was found by LHCb with 40~pb$^{-1}$ of data~\cite{Aaij:2012vn} and, with 1~fb$^{-1}$ of data, they determined~\cite{Aaij:2015gna} it to be 0.2\% of the signal with an overall purity of more than 0.99. At 500 GeV, STAR did not report any same-sign events~\cite{STAR:2011aa} although with a --necessary less clean-- di-electron sample.

Given the likely 
smaller background at energies below 500 GeV, the fact that the muon channel --with smaller background that for RHIC studies-- would  be 
preferred with a much reduced background and the likely strong dependence of any background 
simulation on specific detector performances, we chose not to perform any generator level 
simulation and tend to advocate, in view of past experiments at higher energies that, using 
conventional cuts resulting in tractable acceptances and efficiences (see above), such signals should 
easily be extractable.

\section{Conclusions}
The current planning of future proton colliders necessitates a discussion 
of whether these facilities could also be used in a fixed-target mode. 
There is a long list of physics arguments that supports this case at the LHC.
 We have shown calculations specific to Standard Model bosons at the HE-LHC, SppC and FCC-hh. 
These next 
generation of fixed-target experiments would provide access to high-luminosity 
measurements at unique laboratory energies and momentum transfers. Using a 
bent crystal is a viable option to extract high energy beams of protons and
perform --in a parasitic mode-- fixed-target experiment at 
$\sqrt{s} \simeq 170-300$~GeV with 
annual luminosities on the order of tens of inverse femtobarn, \ie\ with 
high enough rates to produce a significant amount 
of Standard Model bosons. Although it offers a priori smaller luminosities, 
an internal gas target is an option which probably 
requires less civil engineering.


\section*{Acknowledgements}

This work is supported in part by the CNRS via the grants PICS-06149 Torino-IPNO \& PEPS4AFTER2.




\section*{References}

\begin{thebibliography}{99}

 \small



 \bibitem{HELHC:2011}
   R.~Assmann {\it et al.}, 
 CERN-ATS-2010-177.


 \bibitem{FCC-hh:2014}
   A.~Ball {\it et al.},  
 FCC-1401101315-DSC.

\bibitem{SppC:2013}
   Q.~Qin {\it et al.},
 Proceedings of PAC2013, 523 (2013).


\bibitem{Arduini:1997nb}
  G.~Arduini, C.~Biino, M.~Clement, K.~Cornelis, N.~Doble, K.~Elsener, G.~Ferioli and G.~Fidecaro {\it et al.},
  Phys.\ Rev.\ Lett.\  {\bf 79} (1997) 4182.

\bibitem{Elsener:1995hm}
  K.~Elsener, G.~Fidecaro, M.~Gyr, W.~Herr, J.~Klem, U.~I.~Uggerhoj, E.~Weisse and S.~P.~Moller {\it et al.},
  Nucl.\ Instrum.\ Meth.\ B {\bf 119} (1996) 215.


\bibitem{Baurichter:2000wk}
  A.~Baurichter, C.~Biino, M.~Clement, N.~Doble, K.~Elsener, G.~Fidecaro, A.~Freund and L.~Gatignon {\it et al.},
  Nucl.\ Instrum.\ Meth.\ B {\bf 164-165} (2000) 27.

\bibitem{Biryukov:2005gr}
  V.~M.~Biryukov, V.~N.~Chepegin, Y.~A.~Chesnokov, V.~Guidi and W.~Scandale,
  Nucl.\ Instrum.\ Meth.\ B {\bf 234} (2005) 23.

\bibitem{Scandale:2011zz}
  W.~Scandale,  G.~Arduini, R.~Assmann, C.~Bracco, F.~Cerutti, J.~Christiansen, S.~Gilardoni, E.~Laface {\it et al.},
  Phys.\ Lett.\  B {\bf 703 } (2011)  547.
 

\bibitem{Wienands:2015hda}
  U.~Wienands, T.~W.~Markiewicz, J.~Nelson, R.~J.~Noble, J.~L.~Turner, U.~I.~Uggerh\o j, T.~N.~Wistisen and E.~Bagli {\it et al.},
  Phys.\ Rev.\ Lett.\  {\bf 114} (2015) 7,  074801.
 

\bibitem{Uggerhoj:2005xz}
  E.~Uggerh\o j, U.~I.~Uggerh\o j,
  Nucl.\ Instrum.\ Meth.\  B {\bf 234} (2005) 31.


\bibitem{LUA9} W.~Scandale, {\it et al.} [LUA9], CERN-LHCC-2011-007, 2011.

\bibitem{LHCC107} LHC Committee, minutes of the 107th meeting, CERN/LHCC 2011-010


\bibitem{LMC-173} W. Scandale, talk at 173th meeting of the LHC Machine Committee,  \href{https://espace.cern.ch/lhc-machine-committee/Minutes/1/lmc_173.pdf}{slides}


 \bibitem{Brodsky:2012vg}
  S.~J.~Brodsky, F.~Fleuret, C.~Hadjidakis, J.P.~Lansberg,
Phys.\ Rept.\ {\bf 522} (2013) 239.


\bibitem{Lansberg:2012kf}
  J.~P.~Lansberg, S.~J.~Brodsky, F.~Fleuret and C.~Hadjidakis,
  Few Body Syst.\  {\bf 53} (2012) 11
  [arXiv:1204.5793 [hep-ph]].


\bibitem{Lorce:2012rn}
  C.~Lorce, M.~Anselmino, R.~Arnaldi, S.~J.~Brodsky, V.~Chambert, J.~P.~Didelez, B.~Genolini and E.~G.~Ferreiro {\it et al.},
  AIP Conf.\ Proc.\  {\bf 1523} (2012) 149
  [arXiv:1212.0425 [hep-ex]].


\bibitem{Rakotozafindrabe:2013au}
  A.~Rakotozafindrabe, M.~Anselmino, R.~Arnaldi, S.~J.~Brodsky, V.~Chambert, J.~P.~Didelez, E.~G.~Ferreiro and F.~Fleuret {\it et al.},
  Phys.\ Part.\ Nucl.\  {\bf 45} (2014) 336
  [arXiv:1301.5739 [hep-ex]].

\bibitem{Lansberg:2014myg}
  J.~P.~Lansberg, M.~Anselmino, R.~Arnaldi, S.~J.~Brodsky, V.~Chambert, W.~den Dunnen, J.~P.~Didelez and B.~Genolini {\it et al.},
  EPJ Web Conf.\  {\bf 85} (2015) 02038
  [arXiv:1410.1962 [hep-ex]].



\bibitem{Massacrier:2015nsm}
  L.~Massacrier, M.~Anselmino, R.~Arnaldi, S.~J.~Brodsky, V.~Chambert, W.~d.~Dunnen, J.~P.~Didelez and B.~Genolini {\it et al.},
  arXiv:1502.00984 [nucl-ex].


\bibitem{Rakotozafindrabe:2012ei}
  A.~Rakotozafindrabe, R.~Arnaldi, S.~J.~Brodsky, V.~Chambert, J.~P.~Didelez, B.~Genolini, E.~G.~Ferreiro and F.~Fleuret {\it et al.},
  Nucl.\ Phys.\ A {\bf 904-905} (2013) 957c
  [arXiv:1211.1294 [nucl-ex]].


\bibitem{Lansberg:2013wpx}
  J.~P.~Lansberg, R.~Arnaldi, S.~J.~Brodsky, V.~Chambert, J.~P.~Didelez, B.~Genolini, E.~G.~Ferreiro and F.~Fleuret {\it et al.},
  EPJ Web Conf.\  {\bf 66} (2014) 11023
  [arXiv:1308.5806 [hep-ex]].

\bibitem{Massacrier:2015qba}
  L.~Massacrier, B.~Trzeciak, F.~Fleuret, C.~Hadjidakis, D.~Kikola, J.~P.~Lansberg and H.-S.~Shao,
  arXiv:1504.05145 [hep-ex].

\bibitem{Adare:2010xa}
  A.~Adare {\it et al.}  [PHENIX Collaboration],
  Phys.\ Rev.\ Lett.\  {\bf 106} (2011) 062001
  [arXiv:1009.0505 [hep-ex]].

\bibitem{Aggarwal:2010vc}
  M.~M.~Aggarwal {\it et al.}  [STAR Collaboration],
  Phys.\ Rev.\ Lett.\  {\bf 106} (2011) 062002
  [arXiv:1009.0326 [hep-ex]].




\bibitem{Aad:2012tfa}
  G.~Aad {\it et al.}  [ATLAS Collaboration],
  Phys.\ Lett.\ B {\bf 716} (2012) 1
  [arXiv:1207.7214 [hep-ex]].

\bibitem{Chatrchyan:2012ufa}
  S.~Chatrchyan {\it et al.}  [CMS Collaboration],
  Phys.\ Lett.\ B {\bf 716} (2012) 30
  [arXiv:1207.7235 [hep-ex]].

 \bibitem{LHCb:SMOG} C. Barschel, PhD thesis, RWTH Aachen U., Germany, CERN-THESIS-2013-301 (2014)

\bibitem{FerroLuzzi:2005em}
   M.~Ferro-Luzzi,
   {\it Nucl.\ Instrum.\ Meth.\ A} {\bf 553} (2005) 388.
   
\bibitem{LHCb:2014}
R. Aaij \textit{et al.}, [LHCb collaboration], 2014 \textit{JINST} 9 P12005.   

\bibitem{Baldit:1994jk}
  A.~Baldit {\it et al.}  [NA51 Collaboration],
  Phys.\ Lett.\ B {\bf 332} (1994) 244.


\bibitem{Zhu:2007aa}
  L.~Y.~Zhu {\it et al.}  [NuSea Collaboration],
  Phys.\ Rev.\ Lett.\  {\bf 100} (2008) 062301
  [arXiv:0710.2344 [hep-ex]].

\bibitem{Gautheron:2010wva}
  F.~Gautheron {\it et al.}  [COMPASS Collaboration],
  CERN-SPSC-2010-014.
\bibitem{Chiosso:2013ila}
  M.~Chiosso [COMPASS Collaboration],
  Phys.\ Part.\ Nucl.\  {\bf 44} (2013) 6,  882.

\bibitem{Mukherjee:2006uu}
  A.~Mukherjee and W.~Vogelsang,
  Phys.\ Rev.\ D {\bf 73} (2006) 074005
  [hep-ph/0601162].


\bibitem{Campbell:2000bg}
  J.~M.~Campbell and R.~K.~Ellis,
  Phys.\ Rev.\ D {\bf 62} (2000) 114012
  [hep-ph/0006304].


\bibitem{Lai:2010vv}
  H.~L.~Lai, M.~Guzzi, J.~Huston, Z.~Li, P.~M.~Nadolsky, J.~Pumplin and C.-P.~Yuan,
  Phys.\ Rev.\ D {\bf 82} (2010) 074024
  [arXiv:1007.2241 [hep-ph]].




\bibitem{Alves:2008zz}
  A.~A.~Alves, Jr. {\it et al.}  [LHCb Collaboration],
  JINST {\bf 3} (2008) S08005.


\bibitem{Aaij:2012vn}
  R.~Aaij {\it et al.}  [LHCb Collaboration],
  JHEP {\bf 1206} (2012) 058
  [arXiv:1204.1620 [hep-ex]].

\bibitem{Aaij:2015vua}
  R.~Aaij {\it et al.}  [LHCb Collaboration],
  JHEP {\bf 1505} (2015) 109
  [arXiv:1503.00963 [hep-ex]].

\bibitem{Alwall:2007st}
  J.~Alwall {\it et al.},
  JHEP {\bf 0709} (2007) 028
  [arXiv:0706.2334 [hep-ph]].

\bibitem{FONLL}
  M.~Cacciari, M.~Greco and P.~Nason,
  JHEP {\bf 9805} (1998) 007
  [arXiv:hep-ph/9803400];
  M.~Cacciari, S.~Frixione and P.~Nason,
  JHEP {\bf 0103} (2001) 006
  [arXiv:hep-ph/0102134].
  M.~Cacciari, P.~Nason and R.~Vogt,
  Phys. Rev. Lett.  {\bf 95} (2005) 122001
  [arXiv:hep-ph/0502203].



\bibitem{Nadolsky:2008zw}
  P.~M.~Nadolsky, H.~L.~Lai, Q.~H.~Cao, J.~Huston, J.~Pumplin, D.~Stump, W.~K.~Tung and C.-P.~Yuan,
  Phys.\ Rev.\ D {\bf 78} (2008) 013004
  [arXiv:0802.0007 [hep-ph]].



\bibitem{STAR:2011aa}
  L.~Adamczyk {\it et al.}  [STAR Collaboration],
  Phys.\ Rev.\ D {\bf 85} (2012) 092010
  [arXiv:1112.2980 [hep-ex]].




\bibitem{Adamczyk:2014xyw}
  L.~Adamczyk {\it et al.}  [STAR Collaboration],
  Phys.\ Rev.\ Lett.\  {\bf 113} (2014) 072301
  [arXiv:1404.6880 [nucl-ex]].

\bibitem{Adare:2015gsd}
  A.~Adare {\it et al.}  [PHENIX Collaboration],
  arXiv:1504.07451 [hep-ex].

\bibitem{Chatrchyan:2012nt}
  S.~Chatrchyan {\it et al.}  [CMS Collaboration],
  Phys.\ Lett.\ B {\bf 715} (2012) 66
  [arXiv:1205.6334 [nucl-ex]].


\bibitem{Aaij:2015gna}
  R.~Aaij {\it et al.} [LHCb Collaboration],
  arXiv:1505.07024 [hep-ex].

\expandafter\ifx\csname natexlab\endcsname\relax\def\natexlab#1{#1}\fi
\expandafter\ifx\csname bibnamefont\endcsname\relax
  \def\bibnamefont#1{#1}\fi
\expandafter\ifx\csname bibfnamefont\endcsname\relax
  \def\bibfnamefont#1{#1}\fi
\expandafter\ifx\csname citenamefont\endcsname\relax
  \def\citenamefont#1{#1}\fi
\expandafter\ifx\csname url\endcsname\relax
  \def\url#1{\texttt{#1}}\fi
\expandafter\ifx\csname urlprefix\endcsname\relax\def\urlprefix{URL }\fi
\providecommand{\bibinfo}[2]{#2}
\providecommand{\eprint}[2][]{\url{#2}}

\end{thebibliography}
\end{document}